\documentclass[letter]{aa}

\bibpunct{(}{)}{;}{a}{}{,}

\usepackage{graphicx}
\usepackage[varg]{txfonts}
\def\d{$^\circ$}
\def\m{$^\prime$}
\def\s{$^{\prime\prime}$}
\def\hh{$^{\mathrm h}$}
\def\mm{$^{\mathrm m}$}
\def\ss{$^{\mathrm s}$}
\def\cm3{cm$^{-3}$}

\def\2{$^{12}$CO}
\def\3{$^{13}$CO}

\def\msol{M$_\odot$}

\begin{document}

\title{Unveiling the circumstellar environment towards a massive young stellar object} 
\subtitle{}

\author {S. Paron \inst{1,2}
\and C. Fari\~{n}a \inst{3}
\and M. E. Ortega\inst{1}
}

\institute{Instituto de Astronom\'{\i}a y F\'{\i}sica del Espacio (IAFE),
             CC 67, Suc. 28, 1428 Buenos Aires, Argentina \\
             \email{sparon@iafe.uba.ar, mortega@iafe.uba.ar} 
\and FADU and CBC, Universidad de Buenos Aires, Argentina 
\and Isaac Newton Group of Telescopes, E-38700, La Palma, Spain \\
              \email{cf@ing.iac.es}      
}
 
\offprints{S. Paron}

   \date{Received <date>; Accepted <date>}

\abstract{}{As a continuation of a previous work, in which we found strong evidence of massive molecular outflows towards
a massive star forming site, we present a new study of this region based on very high angular resolution observations with the aim 
of discovering the outflow driven mechanism.}{Using near-IR data
acquired with Gemini-NIRI at the broad {\it H}- and {\it Ks}-bands, we study a region of 22\s$\times$22\s around the UCH{\sc ii} region G045.47$+$0.05,
a massive star forming site at the distance of about 8 kpc.
To image the source with the highest spatial resolution possible we employed the adaptative optic system ALTAIR, achieving an angular 
resolution of about 0\farcs15.}{We discovered a cone-like shape nebula with an opening angle of about 90\d~extending eastwards the IR source 
2MASS J19142564+1109283, a very likely MYSO. This morphology suggests a cavity that was cleared in the
circumstellar material and its emission may arise from scattered continuum light, warm dust, and likely emission 
lines from shock-excited gas. 
The nebula, presenting arc-like features, is connected with the IR source through a jet-like structure, which is aligned with the 
blue shifted CO outflow found in a previous study. The near-IR structure lies $\sim$3\s~north of
the radio continuum emission, revealing that it is not spatially coincident with the  UCH{\sc ii} region.
The observed morphology and structure of the near-IR nebula strongly suggest the presence of a precessing jet. In this
study we have resolved the circumstellar ambient (in scale of a thousand A.U.) of a distant MYSO, indeed one of the farthest cases.
}{}


\maketitle

\section{Introduction}

The formation of high-mass stars plays a very important role in the
evolution and dynamics of the Galaxy. Despite the high impact that they
have on the environment, the physical processes involved in their
formation are less understood than those of their low-mass
counterpart.  The birth of a massive star is a violent event,
that produces intense extreme ultraviolet radiation ionizing the
surroundings (e.g. \citealt{hester05,peters10}).  The
formation of a massive star also produces jets and massive molecular
outflows (e.g. \citealt{wu04,vaidya11}) which may contribute to the removal of excess angular
momentum from accreted matter and to disperse infalling circumstellar
envelopes \citep{reip01,prei03}.  Such a phenomenon can take place
even when the object has reached the ultracompact H{\sc ii} (UCH{\sc ii}) region
stage \citep{hunter97,qin08}.  Taking into account that high-mass
stars generally do not form in isolation but in dense clusters and
that the massive star formation sites are located at far distances,
very high angular resolution imaging is required to reveal the
morphologies of their circumstellar ambient.

The UCH{\sc ii} region \object{G045.47$+$0.05} (hereafter G45.47) is located
towards the eastern border of the extended H{\sc ii} region G45L
\citep{paron09} and deeply embedded in a dense and dark molecular
cloud. Figure \ref{presentUC}
displays a three-color {\it Spitzer}-IRAC image of a large field where G45.47 is located 
and the white box indicates its position. The IRAC data were extracted from the GLIMPSE Survey \citep{benja03}.
G45.47 is included in the extensive study of UCH{\sc ii} regions of \citet{wood89}, 
who adopted the distance of 9.7 kpc for it.  According to \citet{paron09}, the H{\sc ii} region G45L is
located at a distance of about 8 kpc, same as the UCH{\sc ii} region complex G45.45+0.06 \citep{kuchar94}, which is
the brightest structure in Fig. \ref{presentUC}. By considering that
all these sources belong to a same complex, \citet{ortega12} adopted
the distance of 8.3 kpc for G45.47. However, following \citet{araya02}, distances down to
6 kpc cannot be excluded for this source. Class II CH$_{3}$OH maser
emission at 6.6 GHz was detected towards G45.47 \citep{caswell95} with
a central velocity of 56 km s$^{-1}$, which is in agreement with the
systemic velocities of the H{\sc ii} region complex and the related
molecular cloud.  It is known that such a maser is radiatively pumped by IR emission
from the warm dust associated with massive young stellar objects
(MYSOs). Indeed, in a previous study, \citet{cesaroni92} using NH$_{3}$
lines observations found evidence of collapse towards G45.47.  Based
on high-angular resolution ($\sim$5\s) molecular line observations,
\citet{wilner96} identified several HCO$^{+}$ J=1--0 clumps suggesting
that G45.47 is in the early stages of forming an OB cluster.  G45.47
is associated with an extended source seen at 4.5 $\mu$m in {\it
Spitzer}-IRAC images, the ``extended green object'' \object{EGO G45.47+0.05}, 
which was cataloged by \citet{cyga08} as a ``likely''
MYSO outflow candidate. Recently, \citet{ortega12} using intermediate
angular resolution observations from several molecular transitions,
characterized the dense molecular clump where G45.47 is embedded and
found strong evidence of outflow activity in the region.

In this study, we report the results obtained from near-IR observations performed with 
Gemini-NIRI with high-angular resolution towards G45.47.

\section{Observations and data reduction}

\begin{figure}[h]
\centering
\includegraphics[width=8cm]{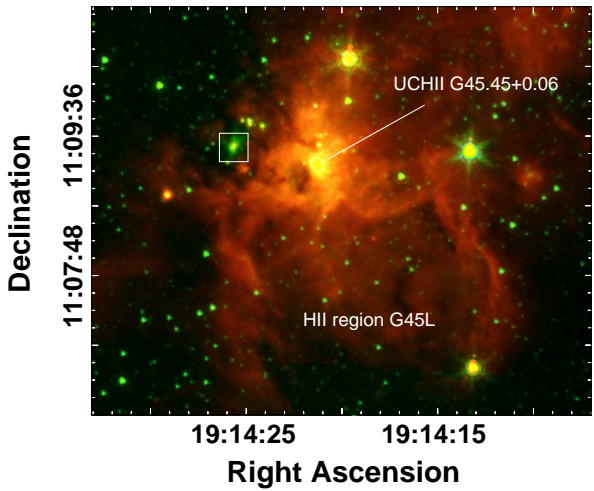}
\caption{Three-color {\it Spitzer}-IRAC image of a large field where the UCH{\sc ii} region G45.47 is located
(8 $\mu$m in red, 4.5 $\mu$m in green, and 3.5 $\mu$m in blue).
The white box indicates the field observed with Gemini-NIRI and G45.47 is the extended green source within it.}
\label{presentUC}
\end{figure}

In this study we analyzed the {\it H} and {\it Ks} broad-band images of the UCH{\sc ii} region G45.47 acquired with NIRI \citep{hodapp03} 
at Gemini-North. 
The observations were carried out during April and June 2012 in queue mode (Program GN-2012A-Q-20). 
NIRI was used with the f/32 camera that provides a plate scale of 0\farcs022 pix$^{-1}$ in a field of view of 22\s$\times$22\s. 
The white box in Fig. \ref{presentUC} represents the field observed with NIRI centered at 
RA $=$ 19\hh 14\mm 25.6\ss, dec. $= +$11\d 09\m 28.5\s, J2000. To image the source with the highest spatial 
resolution possible, NIRI was used together with the Gemini facility adaptive optics (AO) system, ALTAIR \citep{herriot00,boccas06}. For the AO the laser 
guide star system (LGS) was necessary (with a natural guide star only for tip/tilt correction) because G45.47 is 
deeply embedded in a dense molecular clump with high extinction and it was not possible to find a nearby star that fulfill the AO 
brightness requirement in the optical V-band. The available natural tip/tilt star was offset from the UCH{\sc ii} region as well and was 
relatively faint which resulted in a slight elongation of the PSF, noticeable mainly in the {\it H}-band. 
The images in both, {\it H}- and {\it Ks}-bands were performed using a dither pattern 
with offsets to sky fields. The images were reduced using standard procedures for near-infrared (near-IR) imaging provided by the Gemini 
Observatory through the IRAF-NIRI package. The effective seeing was $\sim$0\farcs15  (as measured as the FWHM of point sources in the 
field of the final co-added images). After excluding several individual frames with anomalous electronic noise and with evident 
bad-quality PSFs, the effective exposure times of the final co-added images were 50 s and 120 s for the {\it Ks}- and {\it H}-band, respectively. 
The signal to noise of the extended emission is $\sim$20 for the {\it Ks}-band and $\sim$7 for the {\it H}-band. 
The absolute astrometry in NIRI images was performed using three near-IR stars in the field from the 2MASS Point Source Catalog, 
fortunately the three stars were well distributed in the field allowing a good astrometrical solution.

\section{Results and discussion}

\begin{figure}[h]
\centering
\includegraphics[width=9cm]{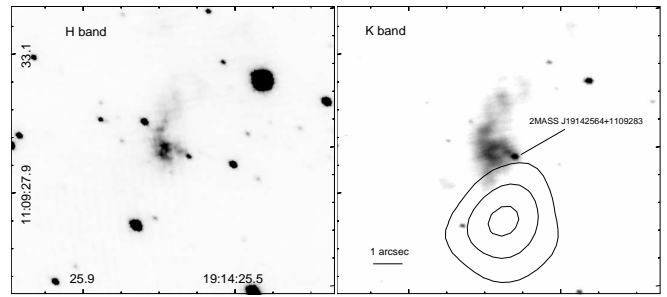}
\caption{Left: {\it H}-band emission towards G45.47. Right: same field displaying the 
{\it Ks}-band emission. The point source 2MASS 19142564+1109283 is indicated and contours 
of the radio continuum emission at 6 cm are included with levels of 5, 20, and 50 mJy beam$^{-1}$.}
\label{hkpan}
\end{figure}

The broad {\it H}-band emission obtained with Gemini-NIRI towards
G45.47 is presented in the left panel of Fig. \ref{hkpan}, while the
right panel displays the broad {\it Ks}-band emission of the same
field.  Both images were slightly smoothed to better display the
diffuse emission that appears at the center
of each panel and eastwards the point source 2MASS
J19142564+1109283. This IR source was suggested to be an early
B-type YSO by \citet{ortega12}. The diffuse emission, better appreciated at the
{\it Ks}-band, shows a nebula with a cone-like shape displaced $\sim$3\s~northwards
the peak of the radio continuum emission at 6 cm, which is displayed with contours. The radio
continuum data was extracted from the RMS Survey \citep{urqu09} and have a synthetized beam size of about 1\farcs7. 
This displacement was previously found by \citet{debuizer05}, who used mid-IR data with a resolution >1\farcs2, and they could not fully 
discard an astrometric problem. Our observations confirm that
the near-IR structure does not coincide with the radio continuum emission, evidencing the complexity of 
this massive star-forming region. Thus, as a first important result, we conclude that the UCH{\sc ii} region G45.47 is 
not directly related to 2MASS J19142564+1109283 and its associated near-IR nebula. 
This is a similar case as found towards the hot molecular core G9.62+0.19, where some UCH{\sc ii} regions, bright in radio continuum,
have not any counterpart in near-IR \citep{linz05}. It seems to be common to find secondary cores slightly separated 
from UCH{\sc ii} regions (e.g. \citealt{beuther07}).

\begin{figure}[h]
\centering
\includegraphics[width=6cm]{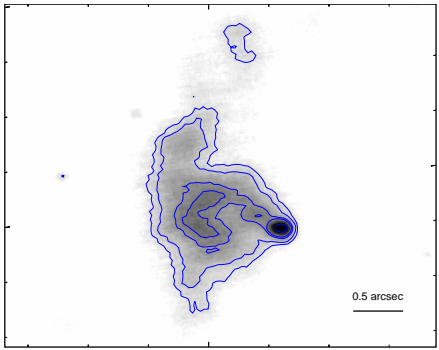}
\caption{Zoom-in of the {\it Ks}-band emission showing the cone-like
  shape nebula. Contours were included to better appreciate the
  morphology. }
\label{kfig}
\end{figure}

\begin{figure}[h]
\centering
\includegraphics[width=6.5cm]{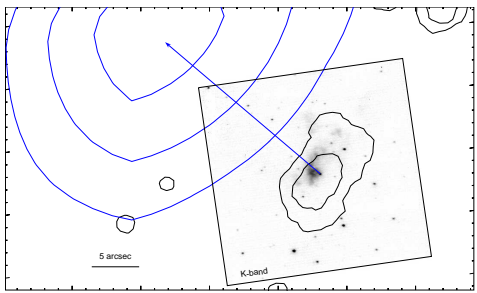}
\caption{The box shows the field observed with Gemini-NIRI and
displays the broad {\it Ks}-band emission. The blue contours are the
blueshifted CO outflow found by \citet{ortega12} and the black ones
represent the 4.5 $\mu$m IRAC emission delimiting the EGO
G45.47+0.05 boundaries.  The angular resolution of the CO emission
is about 20\s.  The blue arrow shows the alignment between the
structure that connects the IR source with the near-IR arc-like
features and the blueshifted CO outflow. }
\label{kcooutf}
\end{figure}

Figure \ref{kfig} shows a zoom-in of the {\it Ks}-band image with 
some emission contours to better display
the morphology.  The cone-like nebula, with an opening angle of about
90\d, presents arc-like features with concave faces pointing to the
IR source. These features seem to be connected to the IR source by a
jet-like structure aligned with the blue shifted CO outflow found by
\citet{ortega12}. Figure \ref{kcooutf} presents the {\it Ks}-band
emission with contours displaying the blueshifted \2 J=3--2 outflow
and the 4.5 $\mu$m IRAC emission. In the image, the blue arrow indicates
the mentioned alignment. Thus, we suggest that the CO outflows are
not related to the UCH{\sc ii} region G45.47. The driving source of the outflows should be
2MASS J19142564+11092832.

The wide-angle ($\sim$90$^{\circ}$) of the cone-like
shape nebulosity pointing to the blue-shifted molecular outflow strongly
suggests that the IR emission arises from a cavity cleared in the
circumstellar material. The diffuse
nebulosity seen at the {\it Ks}-band may be due to a combination of different emitting processes:
continuum emission from the central protostar that is scattered at the inner walls of a
cavity, emission from warm dust, and likely line emission from
shock-excited H$_{2}$, among other emission lines (e.g. \citealt{bik06}). The {\it H}-band emission shown in Fig. \ref{hkpan} (left), which presents
a similar morphology as the {\it Ks}-band image, may also arise from scattered light and warm dust,
and additionally from excited [FeII], tracer of the innermost part of jets that are accelerated 
near the driving source \citep{reip00}.
It is important to note that we do 
not detect any similar feature related to the red-shifted molecular outflow detected by \citet{ortega12}.
This may be due to that the red-shifted component should be much fainter
than the blue-shifted one because it is pointing away from us, which implies a higher extinction. 
Several studies show similar nebulosities related to MYSOs which are associated only with 
the blue-shifted outflow cavity (e.g. \citealt{prei03,kraus2006,weigelt2006}).

An outstanding unsolved problem concerning to jet-driven outflow models is how
highly collimated jets can produce the wide-angle outflow cavities that are usually observed 
in massive protostellar envelopes \citep{mundt1990,konigl2000}.
Two possible scenarios have been proposed to explain this. One of them
proposes a wide-angle structure produced by the winds of the YSO with a density enhancement
along the axis exhibiting a collimated jet-like appearance
\citep{shu1995,reip01}. \citet{weigelt2006} found that the cavity opening angle increases with increasing 
luminosity of the central source, supporting the hypothesis that a
wide-angle stellar wind plays an important role in driving the outflows. 
In the second scenario, the wide-angle cavities are proposed to be excavated by the action of collimated 
precessing jets. \citet{kraus2006} suggest that a precessing jet might explain the difference between the 
outflows widths observed towards low- and high-mass YSOs, adding support to a common
collimation mechanism. According to the authors, the precession of a jet
may be produced in a binary system where the rotational axis of the jet-driving star
is misaligned with the orbital plane of its companion, or may be due to
anisotropic accretion events that alters the angular momentum vector
of the protostar disk. 

\begin{figure}[h]
\centering
\includegraphics[width=6.5cm]{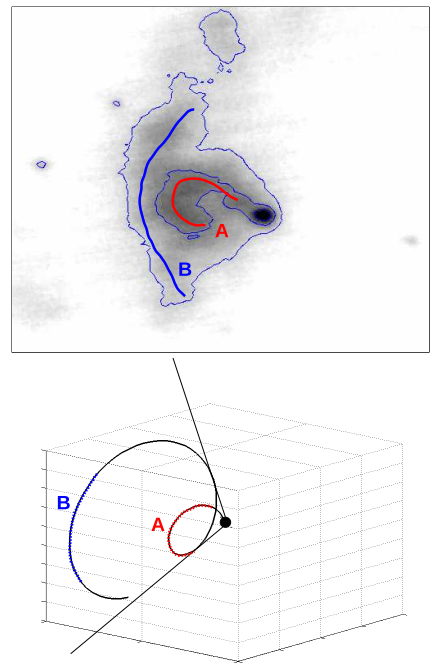}
\caption{Up: {\it Ks}-band emission towards G45.47. Some contours are
superimposed and two arc-like features are remarked in red (feature
A) and blue (feature B). Bottom: sketch of a likely scenario: a
precessing jet coming to us describing an anticlockwise helicoid. The observed
arc-like features A and B are remarked. }
\label{esq}
\end{figure}

The wide-angle stellar wind and precessing jets models predict
different geometries, which can be observationally distinguished. In
the former model, the collimated jet-like structure is seen along the flow axis and the
emission of the nebulosity structures exhibits a highly axial
symmetry. On the other hand, in the precessing jet model, as the jet is free to
move through the cavity, it is expected to observe
asymmetries between the jet position and the emission from the nebulosity.
In our observations, the most prominent arc-like structure (named A in
Fig \ref{esq}-up) is connected to the IR source through an elongated feature
that may trace the most recently ejected material from the
protostar. A clear asymmetry between the axis of the arc-like feature and the
elongated feature is observed, which discards a wide-angle stellar wind scenario.
A less conspicuous arc-like structure (the base of the cone-like structure; named B in Fig. \ref{esq}-up) 
can be observed towards the east. This portion of the nebulosity is clumpy and its concavity points to the 
IR source as in the case of feature A. Thus, the morphology of the observed nebula suggests an spiraling shape produced 
by a precessing jet.
Moreover, the isolated fragment detected towards the north
could be associated with the arc-like feature B, evidencing the
asymmetric nature of the cavity and supporting the interpretation of a
precessing jet.  In Fig. \ref{esq}-bottom we show the arc-like
features named A and B in an sketch of an anticlockwise helicoid, 
with an angle of 50$^{\circ}$~respect to the plane of the sky. 
Following the models presented in \citet{smith2005}, who state that the dominant physical 
structure for all precessing jets is an inward-facing cone, we suggest that the arc-like
morphology observed in our {\it Ks}-band image is consistent with an slow precessing jet.  
\citet{smith2005} show that a fast-precessing jet (precession period $=50$ yr) rapidly disrupts into many bow shocks, while an
slow-precessing jet (precession period $=400$ yr) leads to helical flows generating an spiraling shape nebula.
Following the authors, the observed fragmented helicoid structure in our {\it Ks}-band image can obey to the pulsating 
intrinsic nature of the precessing jet or to inhomogeneities of the circumstellar material.

Binary systems can be found in different stellar evolution stages. 
For example, \citet{connelley2008} found that for Class I YSOs, the binary frequency of systems with separations 
between 100 and 4500~A.U. is about 43\%. \citet{chini12} show a high binarity statistics for O- and B-type stars. 
As previously mentioned, in binary protostars systems with 
misaligned disk, a jet precession is expected due to the tidal interactions between companions (e.g. \citealt{papaloizou1995}). 
Using the relation given in \citet{bate2000}, $P_{prec} \sim 20 \times P_{bin}$, 
where $P_{prec}$ and $P_{bin}$ are the precession and the binary orbital periods, respectively, and assuming a slow 
precessing period of 400~yrs, a $P_{bin} \sim$ 20~yrs is derived. From the generalization of Kepler's third law 
(see eq. 12.15 from \citealt{stahler05}) and considering a relatively large binary system mass of 50 \msol~we 
derive a maximum separation of 27 A.U. between the system components. 
With the angular resolution in our image of 0\farcs15 (1200 A.U. at a distance of 8~kpc) we could not 
spatially resolve any possible binarity under these assumptions. Furthermore, we can not discard the presence of deeply 
embedded sources not detected at near-IR, whose existence could be revealed at longer wavelengths (see \citealt{linz05}). 
Thus, we conclude that binary-induced jet precession may be a plausible scenario.
However, we can not rule out the anisotropic accretion events as the responsible of the jet precession. Spectroscopic observations 
are required to discern between both scenarios. 

\section{Summary}

Using the NIRI instrument at the Gemini North Telescope and the adaptative optic system ALTAIR we studied 
the near-IR emission from the UCH{\sc ii} region G045.47+0.05, a massive star forming site located at a distance 
of about 8 kpc. Achieving an angular resolution of 0\farcs15 we obtained {\it H}- and {\it Ks}-band images 
with a great level of detail, which allowed us to unveil the driven mechanism of the massive molecular outflows
found in a previous study.

We found near-IR diffuse emission with a cone-like shape extending eastwards the IR source 2MASS J19142564+1109283, 
a very likely MYSO. This morphology suggests a cavity that was cleared in the
circumstellar material and its emission may arises from a combination of different emitting processes:
continuum emission from the central protostar that is scattered at the inner walls of a
cavity, emission from warm dust, and likely emission lines from shock-excited gas.
The cone-like nebula, with an opening angle of about
90\d, presents some arc-like features whose concave faces point to the
IR source. These features seem to be connected to the IR source by a
jet-like structure aligned with the blue shifted CO outflow found in a previous study.
The near-IR structure lies $\sim$3\s~north of
the radio continuum emission, revealing that it is not spatially coincident with the UCH{\sc ii} region and 
strongly suggesting that the outflows are 
not generated by it. The driving source of the outflows should be 2MASS J19142564+11092832.
Taking into account the observed asymmetry between the axis of the arc-like features and the
elongated structure that connects them to the IR source, we propose that the spiraling shape of the nebula
is produced by a precessing jet. The observed fragmented helicoid structure may be due to the pulsating intrinsic 
nature of the precessing jet or to inhomogeneities of the circumstellar material.
We propose that the jet precession may be due to tidal interactions in an unresolved binary system or 
to anisotropic accretion events occurring in one MYSO. Spectroscopic observations are required to discern between both scenarios.
In any case, we conclude that we are resolving the circumstellar ambient of a distant  (6-9 kpc) MYSO, indeed 
one of the farthest cases.

\begin{acknowledgements}
We would like to thank the anonymous referee
for her/his extremely helpful comments.
S.P. and M.O. are members of the {\sl Carrera del 
investigador cient\'\i fico} of CONICET, Argentina. 
This work was partially supported by grants awarded by CONICET, ANPCYT and UBA (UBACyT).

\end{acknowledgements}

\bibliographystyle{aa} 
\bibliography{ref} 

\end{document}